\documentstyle[sprocl]{article}

\def\to{\rightarrow}

\def\als{\alpha_s}
\newcommand{\as}{{\ifmmode \alpha_S \else $\alpha_S$ \fi}}

\def\AmS{{\protect\the\textfont2
  A\kern-.1667em\lower.5ex\hbox{M}\kern-.125emS}}
\catcode`@=11
\def\Biggg#1{\hbox{$\left#1\vbox to 22.5\p@{}\right.\n@space$}}
\catcode`@=12

\newcommand\epem{\ifmmode e^+e^- \else $e^+e^-$ \fi}

\newcommand\ms{\ifmmode{\overline{\rm MS}}\else $\overline{\rm MS}$\ \fi}

\newcommand\MSB{\ifmmode{\overline{\rm MS}}\else $\overline{\rm MS}$\ \fi}

\newcommand{\ep}{\epsilon}

\newcommand{\mub}{\ifmmode \mu{\rm b} \else $\mu{\rm b}$ \fi}
\newcommand\alwaysmath[1]{\ifmmode #1 \else $#1$ \fi}

\newcommand{\LQCD}{\ifmmode \Lambda_{\rm QCD} \else $\Lambda_{\rm QCD}$ \fi}
\newcommand{\LMSB}{\ifmmode \Lambda_{\overline{\rm MS}} \else
          $\Lambda_{\overline{\rm MS}}$ \fi}

\def\pp{\ifmmode p\bar{p} \else $p\bar{p}$ \fi}

\def\LMSb{\ifmmode \Lambda_{\rm \overline{MS}} \else
$\Lambda_{\rm \overline{MS}}$ \fi}
\def\abs#1{\left| #1\right|}

\def\I{{\cal I}}

\def\S{{\cal S}}

\def\to{\rightarrow}
\def\d{{\rm d}}

\def\Li{{\rm Li_2}}
\def\del#1{\lower.25em\hbox{\LARGE $\times$}\kern -1em #1 }
\def\st#1{\lower.25em\hbox{$|_{#1}$} }  
\newcommand{\eps}{\epsilon}
\newcommand{\ve}{\varepsilon}
\newcommand\epb{\overline{\epsilon}}
\newcommand{\beq}{\begin{equation}}
\newcommand{\eeq}{\end{equation}}
\newcommand{\beqn}{\begin{eqnarray}}
\newcommand{\eeqn}{\end{eqnarray}}
\newcommand{\beqns}{\begin{eqnarray*}}
\newcommand{\eeqns}{\end{eqnarray*}}
\def\abs#1{\left| #1\right|}

\def\nn{\nonumber}

\def\ms{$\overline{{\rm MS}}$}

\newcommand{\y}{\gamma}

\newcommand{\cg}{c_\Gamma}

\relax

\def\ar#1#2#3{
        {\it Ann. Rev. Nucl. Part. Sci. }{\bf #1} (19#3) #2}

\def\jmp#1#2#3{
        {\it J. Math. Phys. }{\bf #1} (19#3) #2}

\def\np#1#2#3{
        {\it Nucl. Phys. }{\bf #1} (19#3) #2}

\def\pl#1#2#3{
        {\it Phys. Lett. }{\bf #1} (19#3) #2}
\def\pr#1#2#3{
        {\it Phys. Rev. }{\bf #1} (19#3) #2}
\def\prep#1#2#3{
        {\it Phys. Rep. }{\bf #1} (19#3) #2}
\def\prl#1#2#3{
        {\it Phys. Rev. Lett. }{\bf #1} (19#3) #2}

\def\zp#1#2#3{
        {\it Z. Phys. }{\bf #1} (19#3) #2}

\def\st{\scriptstyle}

\def\al{\alpha}
\def\be{\begin{equation}}
\def\ee{\end{equation}}
\def\bea{\begin{eqnarray}}
\def\eea{\end{eqnarray}}
\def\caja{\mathsurround=0pt}
\def\eqalign#1{\,\vcenter{\openup1\jot \caja
        \ialign{\strut \hfil$\displaystyle{##}$&$
        \displaystyle{{}##}$\hfil\crcr#1\crcr}}\,}

\begin{document}
\rightline{ETH-TH-96/05}
\rightline{March, 1996\ \ \ }
\vspace*{2cm}
\title{FROM SCATTERING AMPLITUDES
 TO CROSS SECTIONS IN QCD}

\author{ ZOLTAN KUNSZT }

\address{Institute for Theoretical Physics, ETH,  
\\ CH-8093 Zurich, Switzerland}

% \address{February 29, 1996}

\address{\ \ }
\address{\ \ }
\author{\ \ }
\maketitle\abstracts{{\bf Abstract.\ \ }
I describe how to calculate  cross sections
for hard-scattering processes in high energy collisions
at next to leading order in QCD.   I consider infrared-safe quantities
and  I  assume that the scattering amplitudes
are known   in analytic form up to next-to-leading order. The main
topic is  the description of  the  algorithm for the    
  analytic cancellation of   the 
soft and collinear singularities in the  
loop and bremsstrahlung contributions. The method is systematic
and general. It allows the construction of  
 an analytic expression for     finite next-to-leading
order hard scattering cross sections  suitable  for
numerical evaluation.}

\vspace*{2cm}
\thispagestyle{empty}
\begin{center}
{\it Invited lectures presented at the Theoretical
Advanced Study Institute\\
 in Elementary Particle Physics (TASI 95): 
 \lq QCD and Beyond\rq \\
Boulder, CO, June 4-30, 1995}
\end{center}
\newpage

\section{Introduction}

%QCD~\cite{gellmann} has two fundamental properties:
%{\it asymptotic freedom}~\cite{gross} and {\it colour confinement}.
%Since it is a renormalisable field theory  it  can be studied
%  in weak coupling 
%perturbation theory around the Fock vacuum state of  
%free quarks and gluons in terms of an effective coupling constant.
%We know from the data~\cite{partprop} that the effective coupling constant
%$\as=g_s^2/4\pi$ is about   $0.12$ at $Q=90\gev$ (in the \ms\ sscheme).
% Its value increases with decreasing the scale $Q$ 
% such that 
%at the mass scale of  low lying hadrons it reaches the
%strong coupling regime.  Therefore
% the   perturbative description
% is expected to give   good approximation
%when the  
%   relevant momentum  scale for the given phenomenon is far
%  above the proton mass.  

The application of perturbative QCD~\cite{gross} is not straightforward 
 even for reactions with large momentum transfer $Q$.
In higher order corrections soft and collinear momentum
regions may lead to large contributions of
 order 
$(\als \log(Q/m_q))^n$ where $m_q$ denotes a light quark mass.
 These terms give ${\cal O}( 1)$ corrections which destroy the validity 
 of  the  perturbative treatment.
The  applications
of perturbative QCD are  limited to phenomena where such
terms are either cancelled or can be controlled  with improved treatment
(resummation).
Fortunately, in perturbative QCD the soft and collinear structure
 is relatively 
well understood~\cite{colsoprevs}.
 The main features are summarized by
   fundamental cancellation and factorization theorems 
valid in all orders in perturbation theory.

This lecture will show explicitly how the cancellation and factorization
theorems ``work'' in next-to-leading order applications.
I describe  how to combine  virtual and
real next-to-leading order infrared singular amplitudes into
finite physical cross sections.
 The method of calculating
 amplitudes in leading and next-to-leading order 
 have been described 
in great detail by Lance Dixon~\cite{lancetasi}.
 He also discussed 
the asymptotic properties of amplitudes  in the soft and collinear regions
as well as the soft and collinear singular terms appearing in one-loop amplitudes. 
I  describe the singularities as they appear in various
cross section contributions (loop contributions and bremsstrahlung-contributions).
I consider  their universal features and show explicitly that they cancel for
infrared safe quantities. The examples  will always be  chosen
from the  physics of jet-production.
In order to build a numerical program for efficiently   calculating
 hard scattering  cross sections in next-to-leading order (NLO)
accuracy one should use  a well defined  algorithm
for the analytic  cancellation of the soft and collinear singularities.
Several methods are used in the literature. I shall focus on the
subtraction method.

\section{Cancellation and factorization  theorems} 
\subsection{ KLN cancellation theorem}

In simple inclusive reactions, 
such as  the total cross section of $\epem$ annihilation into
quarks and gluons,
 the soft and collinear contributions
cancel~\cite{kinoshitalee}. That is a consequence of the KLN theorem.
 In the simplest examples,  only one high momentum transfer scale 
is relevant,   the effective coupling  becomes
 small and   
the cross section can reliably be calculated in power series
of the effective coupling up to small power 
corrections
\be
R = \frac {\sigma(e^+e^- \to {\rm hadrons})}{\sigma (e^+e^- \to \mu^+
\mu^-)} = \overbrace{(1 + \frac {\alpha_s}{\pi}+...)}^{\bar R} 3\sum_2
e_{q}^{2}
\ee
where 
\be
\bar R = 1 + \frac{\alpha_s(\mu)}{\pi} + \biggl(\frac
{\alpha_s(\mu)}{\pi} \biggr)^2 \bigl[ \pi b_0 \ln \frac{\mu^2}{s} + B_2
\bigr] + ...
\ee
$\as$ is the running coupling constant, 
 $\mu$ is the renormalization scale, $B_2$ is a known constant
given by the NNLO calculation~\cite{nnloepem}  and $b_0$ is the first coefficient
in the beta-function 
\be
b_0=\frac {11 - \frac {2}{3} n_f}{6 \pi}
\ee
where $n_f $ denotes the number of quark flavours.
The  truncated series is $\mu$-dependent but the $\mu$ dependence is 
${\cal O}(\as^{(n+1)})$ if the cross section is calculated to 
${\cal O}( \as^{n})$.

 The KLN theorem remains
valid also for integrating over final states in a limited phase space
region, as is the case of jet production.
The Sterman-Weinberg two-jet cross section~\cite{stermanweinberg}
 is defined by requiring
that all the final state partons are within a back-to-back cone
 of size $\delta$  
provided their energy is less than $\epsilon \sqrt{s}$.
At NLO
\be\eqalign{
\sigma_{\rm 2jet} &= \sigma_{\rm SW} (s,\varepsilon,\delta)
 \cr 
&  = \sigma_{\rm tot} - \sigma_{q \bar q
g}^{(1)} ({\rm all}\  E > \varepsilon \sqrt{s}, 
{\rm all}\ \theta_{ij} > \delta)
\cr &=
\sigma_0 \biggl[ 1 - \frac {4\alpha s}{3\pi}\bigl(4 \ln 2 \varepsilon
\ln \delta + 3 \ln \delta - 5/2 + \pi^2/3\bigr)\biggr]
}
\ee
where $\sigma_0=4\pi\al^2/3s$.
The cancellation theorem here is hidden in the calculation
of the total annihilation cross section.

\subsection{Factorization theorem}

The initial state collinear singularities, 
which in general do not cancel,  
are universal and  process-independent in all 
orders in perturbation theory~\cite{colsoprevs}. 
Therefore they can be cancelled 
by universal collinear counter terms generated
by the \lq renormalization\rq   of the incoming parton densities. The rule
for defining the finite part of this counter term is fixed
by the factorization scheme. As in the case of ultraviolet 
renormalization~\cite{collinstasi},
 the physics is unchanged under a change of the
factorization scheme, provided the parton densities are also changed
suitably. This feature is expressed by the Altarelli-Parisi 
evolution equation of  parton densities. The collinear
subtraction terms
define  the kernels of the evolution equations.

The  differential
cross section for hadron collisions can be written 
as
\beq
d\sigma_{AB}(p_A,p_B)=\sum_{ab}\int dx_1 dx_2 f_{a/A}(x_A)
f_{b/B}(x_B)d\hat{\sigma}_{ab}(x_A p_A,x_B p_B)\,,
\label{factth}
\eeq
where $A$ and $B$ are the incoming hadrons, $p_A$ and $p_B$
their momentum, and the sum runs over all the parton flavours
which give a non-trivial contribution. The quantities
$d\hat{\sigma}_{ab}$ are the {\it subtracted} partonic cross sections,
in which the singularities due to collinear emission
of massless partons from the incoming partons have been cancelled
by some suitable counter terms. 

According to the factorization theorem, 
the subtracted cross section is obtained by adding the collinear
counter terms to 
the unsubtracted cross section. The latter quantity 
 can be directly calculated in perturbative
QCD. 
Due to universality, eq.~(\ref{factth}) applies also when 
the incoming hadrons are formally substituted for partons. 
In this case, we are also able to evaluate the partonic
densities, which at NLO read
\beq
f_{a/d}(x)=\delta_{ad}\delta(1-x)-\frac{\as}{2\pi}
\left(\frac{1}{\epb}P_{a/d}(x,0)-K_{a/d}(x)\right)
+{\cal O}\left(\as^2\right),
\eeq
where $P_{a/d}(x,0)$ are the Altarelli-Parisi kernels in four 
dimensions (since we will usually work in $4-2\ep$ dimensions, the $0$ 
in the argument of $P_{a/d}$ stands for $\ep=0$) and the functions
$K_{a/d}$ depend upon the subtraction scheme in which the calculation
is carried out. For ${\rm \overline{MS}}$, $K_{a/d}\equiv 0$. Writing
the perturbative expansion of the unsubtracted and subtracted partonic 
cross sections at next-to-leading order as
\beq
d\sigma_{ab}=d\sigma_{ab}^{(0)}+d\sigma_{ab}^{(1)}\,,\;\;\;\;
d\hat{\sigma}_{ab}=d\hat{\sigma}_{ab}^{(0)}+d\hat{\sigma}_{ab}^{(1)}\,,
\label{decomposition}
\eeq
where the superscript 0 (1) denotes the leading (next-to-leading)
order contribution, we have
\beqn
d\hat{\sigma}_{ab}^{(0)}(p_1,p_2)&=&d\sigma_{ab}^{(0)}(p_1,p_2)
\\*
d\hat{\sigma}_{ab}^{(1)}(p_1,p_2)&=
&d\sigma_{ab}^{(1)}(p_1,p_2)
+ d\sigma_{ab}^{\rm count}(p_1,p_2)
\label{counterterms}
\eeqn
where
\beqn
 d\sigma_{ab}^{\rm count}(p_1,p_2)&=&
\frac{\as}{2\pi}\sum_d\int dx\left(\frac{1}{\epb}P_{d/a}(x,0)
-K_{d/a}(x)\right)d\sigma_{db}^{(0)}(xp_1,p_2)
\nonumber \\*&&
+\frac{\as}{2\pi}\sum_d\int dx\left(\frac{1}{\epb}P_{d/b}(x,0)
-K_{d/b}(x)\right)d\sigma_{ad}^{(0)}(p_1,xp_2)\,. \nonumber
\\*
\label{counterterms2}
\eeqn
Eq.~(\ref{counterterms2}) defines the 
 collinear counter terms for any finite hard scattering cross section
for processes with quarks and/or gluons in the initial state.
 Notice that in this
equation the Born terms $d\sigma^{(0)}$ are evaluated in
$4-2\ep$ dimensions.

 The KLN theorem and the factorization   theorems, 
constitute the theoretical basis of
the  description of scattering processes of hadrons in 
 perturbative  QCD.
Those physical quantities for which these theorems remain valid
are called infrared safe~\footnote{ In other words
a measurable is infrared safe if
it is  insensitive  to  collinear splittings of  partons 
and/or   emission of  soft gluons.}.
 These theorems    constitute the necessary
consistency condition for the validity of   the fundamental assumption of
the QCD improved parton model. This   assumption is  that for 
 the case of   infrared safe quantities
the perturbative QCD predictions given in terms of partons  are a  good 
approximation to the same quantities measured in terms of
hadrons (up to power corrections  which are small at high momentum
scales).

Provided the higher order corrections at a given order
are larger than the power corrections,   
 one can systematically 
improve the accuracy of the predictions by calculating
  terms of higher and higher order.
Indeed the analysis of  the  experimental
results required the inclusion of  higher order radiative corrections 
for a large number of measured quantities.

\section{Jet cross sections at next-to-leading order}

We consider the cross sections for three jet
production in $\epem$ annihilation and two-jet production
in hadron-hadron  collisions. These cross sections are 
 proportional
to at least one power of $\as$ and  are  studied  
experimentally with  high precision. 
%%%%%%%%%%%%%%%%%%%%%%%%%%%%%%%%%%%%%%%%%%%%%%%%%%%%%%%
%\input epemjet.tex

\subsection{Three-jet production in ${\epem}$ annihilation}
\label{sec:infraredsafe}
%\subsubsection{Virtual and real contributions}
Let us consider the process 
\be
e^-(k_-) + e^+(k_+) = a_1(p_1) + ...\ + a_n(p_n)
\ee
where
$a_i$ denote quarks or gluons and $n=3,4$.
The amplitudes of these processes ${\cal} A^{(n,i)}$  are known  in the tree 
 approximation $(i=0)$ for $n=3,4,5$ partons
and in the one-loop approximation
$(i=1)$  for  $(n=3)$ partons.   Lance Dixon explained how to calculate
these amplitudes  quickly with modern techniques.
It is convenient to consider the squared amplitude divided
by the flux and the spin averaging factor $8 s$ ($s=2k_+k_-$) :
\beq 
\psi^{(n,0)}_{\epem}(\{a_l\}_{1,n};\{p_l\}_{1,n})=\frac{1}{8 s}\
\sum_{\stackrel{\rm colour}{\rm spin}}\abs{{\cal A}^{(n,0)}_{\epem}}\,
\label{bornampdef}
\eeq
and 
\beq
\psi^{(3,1)}_{\epem}(\{a_l\}_{1,3};\{p_l\}_{1,3})=\frac{1}{8 s}\
\sum_{\stackrel{\rm colour}{\rm spin}}\left(
{\cal A}^{(3,0)}_{\epem}\,{\cal A}^{(3,1)*}_{\epem} + 
{\cal A}^{(3,0)*}_{\epem}\,{\cal A}^{(3,1)}_{\epem}\right)
\label{virtual}
\eeq
where 
 $\{v_l\}_{m,m+n}$ is a short-hand notation for  the list
of variables $${v_m, v_{m+1}, ...,v_{m+n}}$$ and we
 indicated the flavour and momentum dependence
only for the $\psi$ functions. 
The one-loop corrections
  to the  production of three
partons 
are given by 
$\psi^{(3,1)}$. They  were   
 calculated first by R.K.~Ellis, Ross
and Terrano (ERT)~{\cite{ert}}. Recently,  a new
derivation using the helicity method,
 where the orientation  with
respect to the beam direction is  not averaged, 
was given by Giele and Glover~\cite{gieleglover}.

The physical cross sections
are obtained  by integrating
 the product of the $\psi$ functions 
and some ``measurement functions'' $S_{X}$ 
over the corresponding phase space volume
\beq
d\sigma^{\rm nlo}=
d\sigma^{\rm Born}
 + 
d\sigma^{\rm virt}
 + 
d\sigma^{\rm real}\,,
\label{bornvirtreal}
\eeq
where 
\beqn
d\sigma^{\rm Born}(s,{X})
&=&\sum_{\{a_l\}_{1,3}}
\psi^{(3,0)}_{\epem}
{\cal S}_{X,3}(\{p_l\}_{1,3};{X}) d\phi_3(\{p_l\}_{1,3})\\
d\sigma^{\rm virt}(s,{X})
&=&\sum_{\{a_l\}_{1,3}}
\psi^{(3,1)}_{\epem}
{\cal S}_{X,3}(\{p_l\}_{1,3};{X}) d\phi_3(\{p_l\}_{1,3})\\
d\sigma^{\rm real}(s,{X})
&=&\sum_{\{a_l\}_{1,4}}
\psi^{(4,0)}_{\epem}
{\cal S}_{X,4}(\{p_l\}_{1,4};{X}) d\phi_4(\{p_l\}_{1,3})
\eeqn
where $X$ stands for the measured physical quantity and  
\be
d\phi_n = \frac{1}{n!}\int \prod_{i=1}^n\frac{d^{d-1}p_i}{(2\pi)^{d-1}}
(2\pi)^{d}\delta^{(d)}(k_+ + k_ - -\sum_{i=1}^n p_i)
\ee
with $d=4-2\ep$. As a result of   complete flavour sum 
the final particles behave as though they were identical; 
this explains the identical particle factor of
 $n!$. All quantities are calculated
in $d=4-2\epsilon$ dimensions.  The singular
terms appear as single or double poles in $\epsilon$.
The singularities, however, cancel in the sum~(\ref{bornvirtreal}).

{\it Infrared safe measurement function.}
The cancellation of the soft and collinear singularities
of the virtual corrections against the singular part of the
real contribution
is independent of  the
form of the measurement functions
provided they 
are insensitive to collinear splitting and soft emission.
  This means that one obtains the same measured result
whether or not a parton splits into two collinear partons and whether or
not one parton emits another parton that carries infinitesimal
momentum.  A physical quantity that is designed to look at
short distance physics should have this property, otherwise it will be
sensitive to the details of parton shower development and
hadronization.  The mathematical requirements for ${\cal S}_{X,3}$ and
${\cal S}_{X,4}$ are that ${\cal S}_{X,4}$ should reduce to 
${\cal S}_{X,3}$ when two of the outgoing partons become collinear:
\be
\label{safeS}
{\cal S}_{X,4}(p_1^\mu,p_2^\mu,(1-\lambda)p_3^\mu,\lambda p_3^\mu)
={\cal S}_{X,3}(p_1^\mu,p_2^\mu,p_3^\mu)
\ee
for $0\le \lambda \le 1$ 
plus similar conditions where the $\lambda$ and $1-\lambda$
factors are inserted to all possible pairs of the momenta in
$S_{X,4}(p_1^\mu,p_2^\mu,p_3^\mu,p_4^\mu)$.

{\it Example of an infrared safe observable.}
 For the sake of illustration I give   the definition of 
 the measurement function for the    shape
variable
 thrust $T$
\be
\S_{T,n} = \delta \bigl(T - \tau_n (p_1^\mu, p_2^\mu, ..., p_n^\mu)\bigr)
\,,\hspace{.5cm} {\rm where}\hspace{.5cm}
\tau_n = \displaystyle\mathop{\max}_{\vec u} 
{{\sum\limits_{i=1}^{n}\abs{\vec p_i \vec u}}\over
{\sum\limits_{i=1}^n \abs{\vec p_i}}}\,.
\ee
Thrust is  well defined  for arbitrary number of
final-state particles. It is easy to check that it satisfies
the conditions of infrared safety formulated with
the help of eq.~(\ref{safeS}).
%%%%%%%%
\begin{table}[htbp]
\begin{center}
{
\begin{tabular}{|lcl|l|}
\hline
Two & particles & $T = 1$ & Discontinuous\\
Three & particles & 1 $> T >$ 2/3 & in particle\\
Four & particles & 1 $> T >$ 1/ $\sqrt{3}$ & multiplicity\\
$\infty$ & particles & 1 $> T >$ 1/2 & \\
\hline
\end{tabular}
}
\vskip 0.1cm
\caption{ Range of thrust for various numbers of partons}
\label{thrusttable}
\end{center}
\end{table}
\noindent

Thrust  measures the sum of the lengths of the longitudinal momenta
of the final particles relative to the thrust axis $\vec{n}$
chosen to maximize the sum. For two-particle final states its value
is 1. Its allowed range  changes with the particle number,
therefore its differential distribution is only well defined after
some smearing\,\footnote{Smearing is required also by hadronization
  effects. The typical width of a Gaussian smearing is 
$\Delta T = m_h/\sqrt{s}\approx 0.02\,.$}. 
Carrying out the phase-space integral for the Born term one gets
\be
\frac{1}{\sigma_0}
 \;  \frac{d \sigma^{\rm Born}}{dT} \; = \; \frac{\alpha_s}{2
\pi} \; \frac{4}{3} \; \biggl[ \frac{2 (3T^2 - 3T + 2)}
{ (1-T)T} \; \ln
\; \left(\frac{2T - 1}{1 - T} \right) - \frac{
3 (3T - 2) (2 - T)}{ (1-T)} \biggr]
\ee

In the classic paper of ERT 
the singular pieces have been evaluated analytically, and it was 
 demonstrated  that indeed for infrared safe quantities they cancel
as  is required by the KLN theorem.
The remaining finite next-to-leading order
cross section is suitable for  
  numerical evaluation. ERT calculated the  distribution 
for the shape variable $C$. 
 But their analytic  result   can be used to calculate 
any infrared safe three-jet-like quantity in next-to-leading order.
It is convenient to give the one-dimensional 
distributions of various three jet
measures generically denoted as 
$ X=t,C...E_J$ in a form that
satisfies the renormalization group equation
\be
{1\over \sigma_0} {{d\sigma}\over{ dX }}
 ={{\as (\mu)}\over {2\pi}} A_X(X)
+\left({\as (\mu )\over\pi}\right)^2 \Bigl[A_X(X)2\pi b_0 \log
(\mu^2/S)+B_X(X)\Bigr] 
\label{xxsec}
\ee
 $A(x)$ and $B(x)$ are scale-independent functions.
Their values are tabulated for many quantities 
in ref.~{\cite{KunsztNason}}.
The next-to-leading order expression is scale independent
up to  ${\cal O}(\as^3)$
\be
{d\over d\mu^2} \left({d\sigma\over dX}\right) = {\cal O}(\as^3)\ .
\label{xrengr}
\ee
The size and sign of the corrections is rather different
for the various jet measures.
In many cases the corrections are substantial $(\approx 30\%)$
 even at LEP energy.
Thanks to the technical development described by Lance Dixon, 
the NLO  calculation  will soon be   available also for
 four-jet production ( $\psi^{(4,1)}_{\epem} $).
%%%%%%%%%%%%%%%%%%%%%%%%%%%%%%%%%%%%%%%

\subsection{   Jet cross section in hadron-hadron collisions }

At the  Tevatron, multijet cross sections are observed up to 
six jets~\cite{huston}. 
The   analysis of the data requires the evaluation
of the  amplitudes of the  parton processes
\beq \label{hhnproc}
a_1(p_1) + a_2(p_2)\to a_3(p_3) + ... + a_n(p_n)
\eeq
where $a_i$ denotes parton flavour labels, $p_i$ are their four-momenta and 
$n$ is  the number of the participating partons.
The data are rather precise for two- and three-jet like
quantities therefore the comparison with the theory
has to be done at next-to-leading order.
For this purpose 
the tree amplitudes  have to be known  for    $n=4,5,6$ while
   in the case of  $n=4,5$ we have
to know also the one-loop amplitudes. 
It is convenient  again to introduce  $\psi$ functions
giving  the squared amplitude divided by the flux and spin averaging
factors 
\bea 
\psi^{(n,0)}(\{a_l\}_{1,n};\{p_l\}_{1,n})
&=&\frac{1}{2 s\,\omega(a_1)\omega(a_2) }\,
\sum_{\stackrel{\rm colour}{\rm spin}}\abs{{\cal A}^{(n,0)}}\,
\label{bornampdefhh}\\
\psi^{(n,1)}(\{a_l\}_{1,n};\{p_l\}_{1,n})
&=&\frac{1}{2 s\,\omega(a_1)\omega(a_2) }\,
\times\nonumber\\ &&
\sum_{\stackrel{\rm colour}{\rm spin}}
\left({\cal A}^{(n,0)}\,{\cal A}^{(n,1)*} +  
{\cal A}^{(n,0)*}\,{\cal A}^{(n,1)}\right)
\label{virtualhh}
\eea
where $s=2p_1p_2$, ${\cal A}^{(n,i)}$ denote the tree- ($i=0$) and 
 one-loop ($i=1$) amplitudes of the 
process~(\ref{hhnproc}) (helicity, flavour
and momentum labels are all suppressed), and
$\omega(a)$ is the number of colour and spin degrees
of freedom for the flavour $a$,  in $4-2\ep$
dimensions
\be
\omega(q)=2N_c\,,\;\;\;\;
\omega(g)=2(1-\ep)(N_c^2-1)\,.
\ee
The hard-scattering cross section is decomposed
into four contributions
\beq
d\hat{\sigma}^{\rm nlo}=
d\sigma^{\rm Born}
 + 
d\sigma^{\rm virt}
 + 
d\sigma^{\rm real}
 + 
d\sigma^{\rm count}\,,
\label{bornvirtrealcount}
\eeq
where 
\beqn
d\sigma^{\rm Born}_{a_1a_2}(p_1,p_2;{X})
&=&\sum_{\{a_l\}_{3,n}}
\psi^{(n,0)}
{\cal S}_{X,n-2}(\{p_l\}_{3,n};{X})\times\nonumber\\ && 
\hspace*{0.5cm}
d\phi_{n-2}(p_1,p_2;\{p_l\}_{3,n})
\label{hhborn} \\
d\sigma^{\rm virt}_{a_1a_2}(p_1,p_2;{X})
&=&\sum_{\{a_l\}_{3,n}}
\psi^{(n,1)}
{\cal S}_{X,n-2}(\{p_l\}_{3,n};{X})
\times\nonumber\\ &&
\hspace*{0.5cm} d\phi_{n-2}(p_1,p_2;\{p_l\}_{3,n})
\label{hhvirt}\\
d\sigma^{\rm real}_{a_1a_2}(p_1,p_2;{X})
&=&\sum_{\{a_l\}_{1,n+1}}
\psi^{(n+1,0)}
{\cal S}_{X,n-1}(\{p_l\}_{3,n+1};{X})
\times\nonumber\\ &&
\hspace*{0.5cm}
 d\phi_{n-1}(p_1,p_2;\{p_l\}_{3,n+1})
\label{hhreal}
\eeqn
where 
$d\phi_{n-2}(p_1,p_2;\{p_l\}_{3,n})$ is  the phase-space volume
for $n-2$ final particles  with total energy defined by
the incoming four-momenta $p_1$ and $p_2$,  
  $X$  is a generic notation of the measured physical
quantity, $\S_{X,n}$ denotes the  measurement functions of $X$
defined in terms of $n$ partons
 and the counter-term cross sections
$ d\sigma_{ab}^{(\rm count)}(p_1,p_2;{X})$ are given 
 by eq.~(\ref{counterterms2}).
The loop corrections $\psi^{(4,1)}$  have been obtained
for the spin-independent case by Ellis and Sexton~\cite{ES};
the spin-dependent one-loop corrections have been obtained 
by the helicity method~\cite{kusitr} and very 
recently   the one-loop
corrections to the five-parton processes $\psi^{(5,1)}$
have also been calculated (see the lecture
of Lance Dixon).
All  four terms contributing to  the hard scattering
 cross section~(\ref{bornvirtrealcount})
have to be calculated in $4-2\ep$ dimensions. The individual
terms have singular $1/\ep$ and $1/\ep^2$ contributions.
The singularities cancel in the sum, provided the
measurement functions satisfy the conditions of infrared
safety formulated in  
subsection~\ref{sec:infraredsafe}\,\footnote{ 
In the case of hadron-hadron collisions the measurement
functions should also fulfil the condition
that 
${\S}_{X,n-1}$ 
should reduce to ${\S}_{X,n-2}$
 when one of the partons becomes parallel to one of the beam
momenta.}.
The hard scattering cross section given by 
eq.~(\ref{bornvirtrealcount}) is  finite
and,  provided  the cancellation of the singular terms is achieved 
analytically, it is suitable for the numerical evaluation 
of  physical cross sections
with the use of  eq.~({\ref{factth}).

%%%%%%%%%%%%%%%%%%

%\input methods.tex

\section{Methods of analytic cancellation of the singularities}

Although eqs.~(\ref{bornvirtreal}),\,(\ref{bornvirtrealcount})
define finite cross-sections, they cannot be  used directly 
for numerical evaluation since the singular terms in the
real contributions are obtained by integrating
 over the soft and collinear
kinematical range. Since the phase space is large
and its boundary due to the presence of arbitrary
measurement functions is complicated, 
analytic evaluation is impossible.
Fortunately, in the soft and collinear
regions the cross-sections and the measurement
functions have a  simple universal behaviour
such that the integration relevant for the calculation
of the singular terms becomes feasible
analytically. This feature can be implemented 
in two basically equivalent methods.

In the first case ( {\it phase space slicing method} )
 one  excludes (slices) from the numerical integration
domain the  singular regions such that  the numerical
integrations becomes well defined. In the excluded regions
 the $\psi$ functions, $\S$ functions and the relevant
phase-space factors can be replaced  
with their limiting values and  the integrals in these regions
are  performed 
analytically. The boundary of the excluded regions
are defined with some small parameters (invariant mass parameter
or some small angle and energy-fraction parameter
as in the case of the Sterman-Weinberg jet cross-sction).
Let us consider as illustration  a one-dimensional problem
with integration domain $0\le x \le 1$ and an integrand which
has a simple pole at $x=0$.
 One slices the integration region into two pieces,
$0<x<\delta$ and $\delta<x<1$.  We choose $\delta \ll 1$, thus allowing
us to use the simple approximation $F(x) \to F(0)$ for $0 < x < \delta$.
This gives
\begin{eqnarray}
I &\sim& \lim_{\epsilon \to 0}
\left\{
 F(0)\ \int_0^\delta { dx \over x}\ x^\epsilon 
\ +\ 
\int_\delta^1 { dx \over x}\ x^\epsilon\  F(x)
\ - \ {1 \over \epsilon}\ F(0)
\right\}
\nonumber\\
&=&
F(0)\ \ln (\delta)
\ +\ 
\int_\delta^1 { dx \over x}\  F(x) \,.
\label{cutting}
\end{eqnarray}
Now the second integral can be performed by normal Monte Carlo
integration. As long as $\delta$ is small, the sum of the
two terms  will be
independent of $\delta$. About the first use of
this method, see refs.~\cite{Owens,FSKS,Greco}. 
 The method was 
further developed using systematically the
universal features of the soft and collinear limits
by Giele, Glover and Kosower~\cite{gieleglover,GiGlKo}.
The actual implementation is not completely straightforward.
The 
 numerical integration over the real contribution 
has to be done with a certain accuracy. Since the integrand near the 
 boundaries of the singular regions  increases steeply  the 
parameters defining the boundaries must not be very small.
Furthermore, the result of the  analytic integration  
 is  approximate since
the integrand is replaced with its 
limiting value.  The result is more and more accurate
as the parameters defining the boundary of the singular
regions become smaller and smaller. Fortunately,  the measured
cross sections   have finite
 accuracy, which
sets the required precision  for the theoretical evaluation.
 In practical applications one compromises over these
conflicting requirements to achieve  the best efficiency.

The other method is called the {\it subtraction method}.
This method takes account of the fact  that  the  
singular behaviour of the real contribution
is just  some simple pole
with well known simple residue.
After subtraction  the 
 numerical integration over the subtracted integrand becomes
convergent.  The quantity  subtracted should be added back on.
  The
subtraction terms, however, are simple,  
the dimensionally regulated singular
integrations can be carried out  analytically.
This can be illustrated again   with  a simple one-dimensional integral. 
 We write
\begin{eqnarray}
I &=& \lim_{\epsilon \to 0}
\left\{
\int_0^1 { dx \over x}\ x^\epsilon\ [ F(x) - F(0) ]
\ +\ F(0) \int_0^1 { dx \over x}\ x^\epsilon
\right\}
\nonumber\\
&=&
\int_0^1 { dx \over x}\ [ F(x) - F(0) ] \ +  \ {1 \over \epsilon}\ F(0).
\label{subtraction}
\end{eqnarray}
The integral can now be performed by Monte Carlo integration.

This method was  
 first used for QCD
calculations by R.K.~Ellis, Ross, and Terrano\cite{ert} and it was
further developed with systematic use of the simplicity
of the soft and collinear limit in ref.\cite{KS}.
More applications can be found in 
refs.~\cite{KunsztNason,EKS,nrm,FiKuSi}.
The disadvantage of this method is that outside the singular region
the
values of the physical parameters at a given phase-space
point will change if we take the limit corresponding to 
the soft or collinear configurations of the subtraction terms.
In the case of Monte Carlo numerical evaluation this can  lead
 to relatively large fluctuations in the binned distributions.
 This  problem can be avoided by using
Gaussian smearing over the bin size~\cite{KS}.

%%%%%%%%%%%%%%%%%%%%%%%%%%%%%%%%%%%%%%
%\input virtjet.tex

\section{Virtual contributions}

In the following I shall consider only the squared matrix
elements of  process~(\ref{hhnproc}).
The virtual contributions to the cross sections
are given be eq.~(\ref{hhvirt}).
We are interested in  the form  of the singular
terms of the functions $\psi^{(n,1)}$. It turns out that they
have a very simple general structure:
\bea
\psi^{(n,1)}(\{a_l\}_{1,n},\{p_l\}_{1,n})&=&
\frac{\as}{2\pi}\left(\frac{4\pi\mu^2}{Q^2}\right)^{\ep} c_{\Gamma}\nonumber\\
&&\hspace{-1cm}
\left\{-\frac{1}{\ep^2}\sum_{l=1}^n C(a_l)
-\frac{1}{\ep}\sum_{l=1}^n \gamma(a_l)\right\}
\psi^{(n,0)}(\{a_l\}_{1,n},\{p_l\}_{1,n})\nonumber\\
&&
\hspace{-1cm} + \ 
\frac{1}{2\ep} 
\sum_{\stackrel{i,j=1}{i\neq j}}^{n}
\ln \left(\frac{2 \ p_i\cdot p_j}{Q^2}\right)
\psi_{ij}^{(n,0)}(\{a_l\}_{1,n},\{p_l\}_{1,n})\nonumber\\
&&
\hspace{-1cm} + \ 
\psi^{(n,1)}_{{\rm NS}}(\{a_l\}_{1,n},\{p_l\}_{1,n})
 \label{psivirtDR} 
\end{eqnarray}
where we introduced the short-hand notation
\begin{equation}
c_\Gamma=\frac{\Gamma^2(1-\epsilon)\Gamma(1+\epsilon)}
{\Gamma(1-2\epsilon)}\,,
\label{cgdefcol}
\end{equation}
$Q^2$ is an auxiliary variable which cancels in the full
expression, $\mu$ is the scale introduced by dimensional regularization,
  $\psi_{ij}^{(n,0)}$ denotes
  the colour-connected  Born squared matrix elements,
  $C(a_l)$  is the colour charge of parton $a_l$ 
and the constant $\y (a_l)$ gives the size of the virtual
contributions
to the diagonal Altarelli-Parisi kernel $P_{a_l/a_l}(\xi)$
\beqn \label{Candgamma}
C(g) &=&  N_c ; \qquad \quad
\gamma(g) = {11N_c - 2 N_{\rm f} \over 6} \;  \;   \;  {\rm for\ \  gluons}\\
C(q) &=& { N_c^2-1 \over 2 N_c}; \quad
\gamma(q) = {3 (N_c^2-1) \over 4 N} \;  \;   \;  \;  {\rm for\ \  quarks}\,,
\eeqn
finally $\psi^{(3,1)}_{{\rm NS}}$ represents the remaining
finite terms.

The derivation of this result is rather 
simple~\cite{gieleglover,kusitr-sing}.
First we observe that in 
axial gauge the collinear singularities come from the self
energy corrections to the external lines \cite{colsoprevs}.
For each helicity    and colour sub-amplituds 
  they are
 proportional to the Born term 
since  the
Altarelli-Parisi  functions, $P_{a/a}(z)$,
for diagonal splitting preserve helicity in the $z\to 1$ limit.
Therefore, the  collinear 
singularities of the one-loop amplitudes have the form
 \begin{equation}
\label{Colsing}
{\cal A}^{\rm loop}_{\rm col}=
-\left(g\over 4\pi\right)^2 
\sum_a^n {\gamma {(a)}\over \epsilon}  
  {\cal A}^{\rm tree}.
\end{equation} 
There is a contribution for every external leg and
the full contribution to $\psi^{(n,1)}$ is easily obtained
using eq.~(\ref{hhvirt}). 

  The structure of multiple soft
emission from hard processes in QED was investigated by Grammer and Yennie
\cite{Gam73}. They have shown that the energetic electrons
participating in a hard process receive an eikonal phase
factor. In quantum chromodynamics, the situation is very similar
except, that the eikonal factor is a matrix  equal
to the path ordered product of the matrix-valued gluon field
\cite{ColSop81,Cia81,March88}.

For one soft gluon, the main result is very simple:
it states that the singular contributions
come from   configurations where the soft gluons are attached
to the external legs of the graphs. Therefore, 
the soft contribution can easily be calculated in terms
of the Born amplitude. 
The insertion of a soft gluon that connects the external legs $i$ and
$j$ has a twofold effect.
\noindent
First, after carrying  out the loop integral and dropping singular
terms corresponding to collinear configurations, we  pick up
 the same eikonal factor as in QED~\footnote{Using unitarity this form of the soft factor can be confirmed
 by integrating the gluon momenta over  the bremsstrahlung
eikonal factor, as we shall see in the next section.}
\begin{equation}
E_{ij}=
- \left( {g\over 4\pi}\right)^2 c_{\Gamma}
\frac{1}{\epsilon^2}\left(-{ \mu^2\over s_{ij}  }\right)^{\epsilon}.
\label{softsin}
\end{equation}
Secondly, the remaining part of the amplitude
is the same as  the  Born amplitude except that it gets rotated in the
colour space by
 the insertion
of the colour matrices appearing in the two vertices of the
soft line connecting 
 the hard lines  $i$ and $j$, therefore we have the replacement
\begin{equation}
{\cal A}^{(n,0)}_{c_1 c_2..c_n} \to
E_{i,j}\sum_{a, c_i', c_j'}
t^{a}_{ c_i  c_i'}
t^{a}_{ c_j  c_j'}
{\cal A}^{(n,0)}_{c_1... c_i'...c_j'...c_n}
\end{equation}
where  $c_i$'s denote colour indices for the external partons
 and
$t^{a}_{c_i c_i'}$ is the SU(3) generator matrix for the colour
representation of line $i$, that is  $t^{a}_{ij}$ is
$(1/2) \lambda_{ij}^a$        for an outgoing quark,
$-(1/2) \lambda_{ij}^{*\, a}$ for an outgoing antiquark, and
$i f_{aij}$                   for an outgoing  gluon.
For an incoming parton,  the same formula can be used as long
as we use the conjugate colour representations,
$-(1/2) \lambda_{ij}^{*\, a}$ for a quark,
$(1/2) \lambda_{ij}^a$        for an antiquark, and
$i f_{aij}$                   for a gluon.
Finally, using the definition of $\psi^{(n,1)}$ 
we 
obtain 
\bea
\psi^{(n,0)}_{ij}(\{a_l\}_{1,n},\{p_l\}_{1,n})&=&
 \frac{1}{4 p_i\cdot p_j \omega (a_1)\omega (a_2)}\nonumber \\
&&\hspace{-3cm}
2\,{\large {\cal R}eal}
\left(\sum_{spin}\sum_{\{c_l\}_{1,n}}\sum_{a, c_i', c_j'}
t^{a}_{ c_i  c_i'}
t^{a}_{ c_j  c_j'}
{\cal A}^{(n,0)}_{c_1... c_i'...c_j'...c_n} 
{\cal A}^{(n,0)*}_{c_1... c_i'...c_j'...c_n}\right)
\label{psimn}
\eea
This result  of eq.~(\ref{cgdefcol}) 
 is scheme-dependent~\cite{kusitr}. 
In conventional dimensional regularization 
$\psi^{(3,0)}$ and $\psi_{mn}^{(n,0)}$ denote the Born and colour-connected
Born cross section in $4-2\ep$ dimensions and $\psi_{NS}^{(3,1)}$  denotes
the remaining final terms in this scheme.
Finally, we note that to obtain the form of eq.~(\ref{psivirtDR}) 
one should expand the eikonal factor in $\ep$ and
apply the soft-colour identity~\cite{KS}
\beq
\sum_{\stackrel{j=1}{i\neq j}}^{n}\psi^{(n,0)}_{ij}(\{a_l\}_{1,n},\{p_l\}_{1,n}) = 
2\,C(a_i)\,\psi^{(n,0)}(\{a_l\}_{1,n},\{p_l\}_{1,n})\,.
\label{mmnident}
\eeq

%%%%%%%%%%%%%%%%%%%%%%%%%%%%%%%%

%\input hhsoft.tex

\section{Real contributions}

In this section I consider the limiting
behaviour of the real contribution of the process~(\ref{hhnproc})  
 in the soft and collinear limit.
I also give the   local subtraction terms which
render the real contributions integrable over the whole
phase-space region.
The discussion will be detailed on the soft contributions.
In the case of the collinear limits 
I only  summarise their  most salient  features.

\subsection{Kinematics}

The four-momenta of the reaction~(\ref{hhnproc}) 
can be  parameterised, for example,
 in terms of transverse momenta,
rapidities and azimuthal angles
\beqn\nn
&&p_1^\mu \,=\, \frac{\sqrt{s}}{2} (x_1,0,0,x_1)\,, \quad  
p_2^\mu \,=\, \frac{\sqrt{s}}{2} (x_2,0,0,-x_2)\,  \\ \label{fourmom} 
&&p_{i}^\mu \,=\, 
p_{\perp,i} \ ({\rm ch}\ y_i,\cos \phi_i, \sin \phi_i, {\rm sh}\ y_i), 
\; \; i \in \{ 3,...,n \} \,.
\eeqn
From  energy and longitudinal momentum conservation
 we obtain for the momentum fractions
\beq
x_1 \,=\, \frac{1}{\sqrt{s}} \sum_{i=3}^{n} p_i e^{y_i}\,, \quad
x_2 \,=\, \frac{1}{\sqrt{s}} \sum_{i=3}^{n} p_i e^{-y_i}\,.
\eeq
Considering singular limits we shall always assume
the use of some suitable set of independent variables
appropriate for the phase-space integration after imposition 
of four-momentum conservation. 
For the definition of measurable quantities, rapidity and transverse
momentum variables are particularly convenient since they
are boost-invariant~\cite{KS,EKS}.
In the evaluation of
the soft and collinear limit it appears more convenient, however, 
to use energy-angle variables
\beq
p_i^{\mu}=(E_i,
\sin \phi_i\sin \theta_i, \cos \phi_i\sin \theta_i,\cos \theta_i)\,.
\eeq

\subsection{Soft subtraction terms and soft contributions}
The cross section of the real contribution (see eq.~(\ref{hhreal})) 
is constructed from  products of three factors:
function $\psi^{(n+1,0)}$, the measurement function ${\cal S}_{X,n-1}$ and
the phase-space factor $d\phi_{n-1}$.
Considering their soft limits
let us assume that  parton $a_k$ is soft 
($3\le k \le n+1$). 
 Its energy is denoted by $E_k$ and its angular
correlations are controlled by the four-vector $n^{\mu}_k$
defined by the relation
\be
k^{\mu}=E_k n^{\mu}_k=E_k (1,\vec{n}_k)\,.
\ee
The method that has been used in the previous section to calculate
the soft limit of the virtual contribution $\psi^{(n,1)}$
also applies  for the  real emission,  
and we obtain
\bea
\lim_{E_k\to 0}\psi^{(n+1,0)}(\{a_l\}_{1,n+1};\{p_l\}_{1,n+1})
&=& \delta_{ga_k}
\frac{4\pi\as}{E_k^2}
\sum_{i,j;\,i<j}^{[k]}e_{ij}(p_i,p_j,n_k)\times\nonumber\\ &&
\psi^{(n,0)}_{ij}(\{a_l\}^{[k]}_{1,n+1};\{p_l\}^{[k]}_{1,n})
\eea
where the list  $\{v_l\}^{[k]}_{m,m+n}$ denotes the same list
as  $\{v_l\}_{m,m+n}$ but $v_k$ ($m\le k\le m+n$) is left out,
 $\psi^{(n,0)}_{ij}$ is the colour-correlated Born contribution 
defined by eq.~(\ref{psimn}) in the previous section  and $e(p_i,p_j,n_k)$
is the eikonal factor for real emissions:
\be
e(p_i,p_j,n_k) = 
\frac{p_i\cdot p_j}{p_i\cdot n_k~p_j\cdot n_k}.
\label{eij}
\ee
We note that  $e(p_i,p_j,n_k)$ is independent of $E_k$ but is
 dependent on the 
angular variables of the soft line. The colour-correlated Born terms 
$\psi_{ij}$, however, are
 completely independent of the soft momenta.
The soft limit of the 
measurement function is given by the requirement of infrared safety 
\be
\lim_{E_k\to 0}\S_{X,n-1}(\{p_l\}_{3,n+1 };X)=
\S_{X,n-2}(\{p_l\}^{[k]}_{3,n+1};X)
\ee
and  the phase-space factor  behaves as
\beq
\lim_{E_k\to 0}
d\phi_{n-1}(p_1,p_2\to \{p_l\}_{3,n+1})
 = \frac{1}{n-1}d\phi_1[k]d\phi_{n-2}\left(p_1,p_2\to \{p_l\}^{[k]}_{3,n+1}\right)
\eeq
where
\be
d\phi_1([k]) = \mu^{2\ep}
\int \frac{d^{d-1}p_k}{(2\pi)^{d-1}\,2p^0_k} =  \mu^{2\ep}
\int \frac{E^{1-2\ep}_k}{2\,(2\pi)^{(3-2\eps)}} dE_k \int\d\Omega_{3-2\ep}
\ee
is the phase-space integral over parton $a_k$
and the decomposition into energy and angular integrals is also 
shown\,\footnote{
We note that in the soft limit the condition imposed by the
delta function of momentum conservation is different from the original one;
therefore our notation is not completely precise. In the
soft limit some of the components of the momenta $\{p_l\}^{[k]}_{3,n+1}$
will be different from the original ones and 
instead $\{p_l\}^{[k]}_{3,n+1}$ we should  write 
$\lim_{E_k\to 0}\{p_l\}^{[k]}_{3,n+1}$. We tacitly assume that 
in the soft limit $p_l$ denotes its limiting value.
It 
 is uniquely defined after choosing   the independent set of variables 
in the phase space-integral $d\phi_{n-1}$.}.

{\it The local soft subtraction term} for subtracting the soft singular
behaviour in the $E_k\to 0$ limit is
 given by the integrand of the expression below
\bea
d\sigma^{(\rm soft,sub)}_{a_1a_2,k}(p_1,p_2;{X})
&=&-\frac{\delta_{a_kg}}{n-1}\,\frac{\as}{ 2\pi}
\sum_{i,j;\,i<j}^{[k]}\nonumber\\
&&\hspace{-1.5cm} 
\sum_{\{a_l\}^{[k]}_{3,n+1}}
\left[8\pi^2\, e(p_i,p_j,n_k)\,
\frac{1}{E^2_k}\Theta(E_k\ge E_c)d\phi_1[k]\right]
\nonumber \\
&&\psi^{(n,0)}_{ij}(\{a_l\}^{[k]}_{1,n+1};\{p_l\}^{[k]}_{1,n+1})
{\cal S}_{X,n-2}(\{p_l\}^{[k]}_{3,n+1};{X})
\times\nonumber\\ &&
\hspace*{0.5cm}
\frac{1}{(n-2)!}d\phi_{n-2}(p_1,p_2\to \{p_l\}^{[k]}_{3,n+1})
\label{hhsoftsub}
\eeqn
  In eqs. (\ref{hhreal}) and
(\ref{hhsoftsub}), we can use the
same  independent integration variables,
 therefore, by adding  eq.\,(\ref{hhsoftsub})  
to the integrand of (\ref{hhreal}) we subtract its singular
regions defined by the $E_k\to 0$ limit.
Again, to ensure that 
the subtraction does not change
the value of the
original expression  what is subtracted has to be added back on
but in an integrated form where the
singular terms are  calculated analytically. 
This is achieved by  carrying out the 
phase-space integral $d\phi_1[k]$. The expression
in the square bracket in eq.(\ref{hhsoftsub})
defines the  soft integral
\be
\I^{\rm soft}_{ij}(z_{ij},\mu/E_c,\ep)
=8\pi^2\mu^{\ep}\int\frac{d^{d-1}p_k}{2p^0_k(2\pi)^{d-1}}\,
\frac{p_i\cdot p_j}{p_i\cdot p_k\,p_j\dot p_k}  
\Theta(E_k\ge E_c)
\ee
where $E_c$ is a cut-off value on the energy of the particle which
is allowed to be soft.
We also indicated the remaining dependences of  the integral;
 the angular variable
$z_{ab}$ is defined as
 $$z_{ab}=\cos\theta_{ab}=\frac{\vec{p}_a\vec{p}_b}
{\abs{\vec{p}_a}\abs{\vec{p}_b}}.$$

%%%%%%%%%%%%%%%%%%%%%%%%%%%%%%%%%%%%%%%%%%%%%%%
\begin{table}[!htbp]
\begin{center}
{
\begin{tabular}{|c|c|c|}
\hline %--------------------------------
\rule[-1.2ex]{0mm}{6ex} 
  {\it Definition } & {\it Answer }       \\
\hline  % ----------------------------
\rule[-2ex]{0mm}{8ex}
 $ J(\eps)   =
\frac{\Omega_{1-2\ve}}{\Omega_{2-2\ve}}
\int_{-1}^1d\cos\theta
\int _0^{\pi}d\phi\,
\frac{(\sin\theta\sin\phi)^{-2\ve}}{1-\cos \theta} $ 
 &  $ - \frac{1}{\ve} +2\ln 2 $\\
\rule[-2ex]{0mm}{6ex}
$\quad\quad =
\int_{-1}^1 \frac{(1-z^2)^{-\ve}}{1-z} $
&
$ +\ve(\Li(1)-2\ln^2 2)$ 
 \\ 
\hline %--------------------------------
\rule[-2ex]{0mm}{6ex} 
$ I_{\phi}(z_{ij},z_{ik})  = \int_0^{\pi}d\phi_k
\frac{1}{1- z_{ij}z_{ik}
-\sin\theta_{ij}\sin\theta_{ik}\cos\phi_k }
$ &
$\frac{\pi}{\abs{z_{ik}-z_{ij}}}  
$ \\ 
\hline  %---------------------
%\rule[-2ex]{0mm}{6ex} 
% $ l^{R}_{ij}(z_{ij},z_{ik},z_{jk})
%= 
%\frac{1}{2} \biggl(\frac{1-z_{ij}}{(1-z_{ik})(1-z_{jk})}-\frac{1}{1-z_{ik}}-
%\frac{1}{1-z_{jk}}\biggr) 
%$ &{\ }\\
%\hline
\rule[-2ex]{0mm}{6ex} 
 $J^{(0)}
(z_{ij})  = \frac{1}{\pi}\int_{-1}^1dz_{ij}\int_0^{\pi}
d\phi_k\, l^{R}_{ij}$ &
$ \ln\frac{1-z_{ij}}{2} $
\\
\hline
\rule[-2ex]{0mm}{6ex} 
 $J^{(\phi)}
(z_{ij})  = \frac{1}{\pi}\int_{-1}^1dz_{ij}\int_0^{\pi}
d\phi_k\,\ln\sin^2\phi\ l^{R}_{ij}$ &
$\ - \ln 2(1+z_{ij})\ln\frac{1-z_{ij}}{2} $
\\
\hline 
\rule[-2ex]{0mm}{6ex} 
$ J^{(z)}
(z_{ij})  = \frac{1}{\pi}\int_{-1}^1dz_{ij}
\ln\sin^2\theta_{ik}
\int_0^{\pi}
d\phi_k\,  l^{R}_{ij}$ &
$\frac{1}{2}\ln^2 2(1-z_{ij})-2\ln^22$
\\ 
\rule[-2ex]{0mm}{6ex} 
 {} &
$ +\Li(1)-
\Li(\frac{  1-z_{ij}  }{2}) 
$ \\
\hline
\end{tabular}
}
\vskip 0.1cm
\caption{ List of angular integrals}
\label{softintab}
\end{center}
\end{table}
%%%%%%%%%%%%%%%%%%%%%%%%%%%%%%%%%%%%%%%%%%%%%%%%%%%%%%%%%
It is convenient to  decompose the eikonal factor
into a term  which has  one collinear singularity and
terms which are finite 
\beq
\frac{p_i\cdot p_j}{(p_i\cdot p_k)(p_j\cdot p_k)}=\frac{1}{E_k^2}\left[
l(z_{ij},z_{ik},z_{jk})+ (i\leftrightarrow j)\right]
\label{lijdef}
\eeq
where
\beq
l(z_{ij},z_{ik},z_{jk}) =\frac{1}{1-z_{ik}} + l^R(z_{ij},z_{ik},z_{jk})
\label{lsplit}
\eeq
and 
\be
l^R(z_{ij},z_{ik},z_{jk})=
\frac{1}{2}\left[\frac{1-z_{ij}}{(1-z_{ik})(1-z_{jk})}
-\frac{1}{1-z_{ik}}-\frac{1}{1-z_{jk}}\right].
\label{langord}
\ee
The integral over the energy is trivial and 
the angular integrals have to be calculated only  up to order 
$\ep$  so one obtains
\beqn\nonumber
I_{ij}^{\rm \, soft}&=&c_{\Gamma}(\frac{4\pi\mu^2}{E_c^2})^{\ve}
(-\frac{1}{\ve})\\ \label{isoftn}
& &\left[J(\ve)+J^{(0)}(z_{ij})(1-\ve\ln 4)-\ve 
J^{(z)}(z_{ij})+J^{(\phi)}(z_{ij})\right]
\label{jjjint}
\eeqn
where $\cg$ was defined in eq.~(\ref{cgdefcol}).
The integrals 
$J^{(0)}$, $J^{(\phi)}$ and  $J^{(z)}$ are 
listed in Table~(\ref{softintab}).
Inserting their  values into eq.(\ref{jjjint}) and expanding in 
$\ep$ we get both the singular
 and finite parts 
\bea\label{softspliti}
I_{ij}^{\rm\, soft}&=&
I_{ij}^{\rm\, soft,sing}+
I_{ij}^{\rm\, soft,fin}\\ 
 \label{softsingi} 
I_{ij}^{\rm\, soft,sing}&=&c_{\Gamma}
\left({4\pi\mu^2\over Q^2}\right)^\ve
\left[\frac{1}{\ve^2}-
\frac{1}{\ve}\ln \frac{2p_ip_j}{Q^2}+
\frac{1}{\ve}\ln \frac{E_iE_j}{E_c^2}
\right]\,.
\label{isofres}
\eea
The explicit form of the finite part is not interesting
for us here; it can be found in ref.~\cite{FiKuSi}.
We can  use completely covariant 
 notation by replacing 
the energies with the 
covariant expressions
\be
E_i=({\cal P}p_i)/\sqrt{s}\,, \ \  {\rm where}\ \ 
{\cal P}^{\mu}=p_A^{\mu} + p_B^{\mu}=(\sqrt{s},0,0,0)
\label{covEi}
\ee
where $p_A$ and $p_B$ are the four momenta of the incoming hadrons.

If the sum over flavour is carried out 
the result becomes independent of  the label of the soft line. Ttherefore  
every leg in the final state
gives the same contributions and the factor $1/(n-1)$ in
 eq.~(\ref{hhsoftsub})
gets cancelled.
As a result,
{\it the soft contribution} 
of the real leading order process with $n+1$ partons 
can be written  in the  form of the virtual contribution of 
the corresponding $n$-parton process. It
can be obtained form the right hand side of 
eq.~(\ref{hhsoftsub}) by multiplying it 
with $n-1$, changing its sign and inserting the integrated value 
of $\I^{\rm soft}_{ij}$ (eq.~\ref{isofres}):
\bea
d\sigma^{(n+1,{\rm soft})}_{a_1a_2}(p_1,p_2;{X})
&=&\sum_{\{a_l\}_{3,n}}
\psi^{(n+1,{\rm soft})}
{\cal S}_{X,n-2}(\{p_l\}_{3,n};{X})
\times\nonumber\\ &&
\hspace*{0.5cm} d\phi_{n-2}(p_1,p_2;\{p_l\}_{3,n})
\label{hhsoftcont}
\eea
where 
\bea
\psi^{(n+1,{\rm soft})}(\{a_l\}_{1,n},\{p_l\}_{1,n})&=&
\frac{\as}{2\pi}\left(\frac{4\pi\mu^2}{Q^2}\right)^{\ep} 
c_{\Gamma}\nonumber\\
&&\hspace{-2cm}
\left\{\sum_{l=1}^n \frac{C(a_l)}{\ep^2} + 
2 \frac{C(a_l)}{\ep} \ln\frac{E_l}{ E_c} 
\right\}
\psi^{(n,0)}(\{a_l\}_{1,n},\{p_l\}_{1,n})\nonumber\\
&&
\hspace{-3cm} + \ 
\frac{1}{2\ep}
\sum_{\stackrel{i,j=1}{i\neq j}}^{n}
\ln \left(\frac{2 \ p_i\cdot p_j}{Q^2}\right)
\psi_{ij}^{(n,0)}(\{a_l\}_{1,n},\{p_l\}_{1,n})\nonumber\\
&&
\hspace{-1cm} + \ 
\psi^{(n+1,{\rm soft})}_{{\rm NS}}(\{a_l\}_{1,n},\{p_l\}_{1,n})
 \label{psisoftDR} 
\eea
The first term is the soft-collinear singularity
and it cancels the soft-singular terms appearing in the 
virtual corrections. The third term is the soft contribution
proportional to the colour correlated Born-term and again it cancels
the corresponding terms in the virtual contributions.
The second term comes from the collinear singularities of the 
eikonal factors and is  cancelled by the direct singular
collinear contributions. The last term is finite and can be evaluated in
four dimensions.

\subsection{Collinear subtraction terms and collinear contributions}
\newcommand\LC{\stackrel{\sss i\parallel j}{\longrightarrow}}
\newcommand\sss{\scriptscriptstyle}
\newcommand\CA{C_{\sss A}}
\newcommand\DA{D_{\sss A}}
\newcommand\CF{C_{\sss F}}
\newcommand\TF{T_{\sss F}}
In the previous section we constructed the local subtraction term for the
soft singular region. We also demonstrated
that the soft singularities of the
virtual corrections are cancelled 
by  the soft contributions of the real corrections
after  the integrals
over the energy and angular variables of the soft line are carried out.

The
 subtraction and addition  procedure can also be applied to
the singular collinear regions.
 Similarly to the soft case, one finds again 
simple limiting behaviours  for the 
$\psi$-functions, the measurement functions and  the phase space.
We shall discuss only  the collinear limit of the
$\psi$ function.  The reader can find further  details in 
 refs.~\cite{GiGlKo,KS,FiKuSi}.

From  the simple behaviour
of the helicity amplitudes in the collinear limit
we obtain  for the
$\psi$ function the limiting behaviour~\cite{KS,FiKuSi}
\bea
&&\psi^{(n+1,0)}\left(p_1,p_2;\,..,p_i,..,p_j,..\right)\LC
\nonumber \\*&&\phantom{aaaaaaa+}
\frac{4\pi\as}{p_i\cdot p_j}\,P_{a_i S(a_i,a_j)}^<(z)
\psi^{(n,0)}\left(p_1,p_2;\,..,_{\sss P},..\right)
\nonumber \\*&&\phantom{aaaaaaa}
+\frac{4\pi\as}{p_i\cdot p_j}\,Q_{a_i S(a_i,a_j)^\star}(z)
\tilde{\psi}^{(n,0)}\left(p_1,_2;\,..,p_i,..,p_j,..\right),
\label{collimit}
\eeqn
where $z$ denotes the momentum fraction defined through the equations
\beq
p_i=zp_{\sss P}\,,\;\;\;\;k_j=(1-z)p_{\sss P}\,,
\label{zdef}
\eeq 
 $P_{ab}(z)$ denotes the standard Altarelli-Parisi splitting functions
 and  $Q_{ab*}(z)$'s are some new   universal functions which
control the azimuthal angle behaviour of the collinear limit.
Assuming that the collinear particles belong to the final state
\beqn
Q_{gg^{\star}}(z)&=&-4\CA\,z(1-z)\,,
\label{Q1}
\\
Q_{qg^{\star}}(z)&=&4\TF\,z(1-z)\,,
\\
Q_{gq^{\star}}(z)&=&0\,,
\\
Q_{qq^\star}(z)&=&0\,.
\label{Q4}
\eeqn
In these equations, the $^\star$ symbol over the flavour of the 
particle that eventually splits reminds that this particle
is off-shell. In principle, this notation should be extended
also to the Altarelli-Parisi splitting kernels, but at the leading
order $P_{ab^\star}=P_{a^\star b}$, and therefore there is no
need to keep track of the off-shell particle; the $\tilde{ \psi}$
function is constructed from the helicity amplitudes of 
$n$-parton processes but with some linear dependence on the
azimuthal angle of the quasi-collinear  configuration.
The $\tilde{\psi}$ functions therefore are as simple as the
Born-terms. They are important in constructing a correct
local subtraction term. However, upon integrating over
the  azimuthal angle of the collinear momenta
their  contributions vanish and they do not appear 
in the integrated {\it collinear contributions.} 
It is an interesting simplifying 
feature of the five  parton processes that the
$\tilde{\psi}$ functions  vanish identically.

Using  crossing properties of the splitting
functions and of the $\psi$ functions similar relations remain
valid also for initial collinear singularities.

The  remaining  procedure, in principle, is the same what we used in the
soft case
although it is somewhat more tedious. 
The 
properties of the $S$-functions and the phase-space again ensure
that we can easily construct  the local collinear counter terms
and in the collinear contributions we can obtain
the singular terms  analytically.
For further details   the reader 
should  consult the original literature \cite{GiGlKo,KS}.

\noindent {\bf Exercise:} Calculate the inclusive one-jet cross section
$d\sigma/dE_Jd\cos \Theta_J$ for $\epem$ annihilation following the
general algorithm explained above.

\section{Conclusion}
In this lecture I described
the methods and techniques  to get
 differential cross-section formulae 
from NLO singular scattering amplitudes. 
These  are
free from singular terms
  well defined in  all  integration regions
and therefore   suitable for numerical evaluation,
The ingredients of the methods are
  the collinear counter
terms, the 
local subtractions terms,  
 the virtual, the soft and the collinear contributions. 
Their  sum  defines the  
 {\it finite hard scattering cross-section}
\bea
d\hat{\sigma}^{\rm hard}_{a_1a_2}&=&d\sigma^{\rm born}_{a_1a_2} + 
d\sigma^{\rm virt}_{a_1a_2} + 
d\sigma^{\rm soft}_{a_1a_2} +
d\sigma^{\rm( coll, initial)}_{a_1a_2} + 
d\sigma^{\rm (coll, final)}_{a_1a_2}\nonumber\\ &+&
d\sigma^{\rm (coll, counter)}_{a_1a_2}  + 
d\sigma^{\rm (real, subtracted)}_{a_1a_2}\,
\label{hardxsection}
\eea
where  the last term has the kinematics of an $n+1$ parton
process while all the other terms have the kinematics  of an $n$ 
parton process.  In addition, 
the evaluation of  some physical quantity
  requires the explicit construction
of the corresponding measurement functions. We note that 
in the case of jet-production
this 
is a non-trivial exercise (see 
refs.\cite{bkss,es,cdss}). 

We have seen that although the methods are conceptually simple,
their implementation is non-trivial.
Recently,  several papers  attempted to give a   comprehensive
documentation \cite{GiGlKo,KS,FiKuSi,cataniseymour}, which are
 recommended for further reading.

%%%%%%%%%%%%%#############################

\section*{Acknowledgements}
I would like to thank Dave Soper for his significant contribution
to my understanding of the subject of this talk and  Keith Ellis
for reading the manuscript. 
I also thank Dave Soper and K.~T.~ Mahanthappa for their hospitality during my
stay and for  the organization of a great 
summer-school.

\end{document}